\documentclass[aps,showpacs,prb,twocolumn,footinbib,superscriptaddress]{revtex4-1}

\usepackage{epsfig}
\usepackage{amsfonts}
\usepackage{amsmath}
\usepackage{dsfont}
\usepackage{txfonts}
\usepackage{xcolor}

\usepackage[utf8]{inputenc}
\usepackage[colorlinks,bookmarks=false,citecolor=blue,linkcolor=blue,urlcolor=blue]{hyperref}
\usepackage{soul}

\begin{document}
	
	\title{Non-monotonic response and light-cone freezing in gapless-to-(partially) gapped quantum quenches of
		fermionic systems}
	\author{S. Porta}
	\affiliation{Dipartimento di Fisica, Universit\`a di Genova, 16146 Genova, Italy}
	\affiliation{SPIN-CNR, 16146 Genova, Italy}
	\author{F. M. Gambetta}
	\affiliation{Dipartimento di Fisica, Universit\`a di Genova, 16146 Genova, Italy}
	\affiliation{SPIN-CNR, 16146 Genova, Italy}
	\author{N. Traverso Ziani}
	\affiliation{Institute for Theoretical Physics and Astrophysics, University of W\"{u}rzburg, 97074 W\"{u}rzburg, Germany}
	\author{D. M. Kennes}
	\affiliation{{Department of Physics, Columbia University, New York, NY 10027, USA}}
	\author{M. Sassetti}
	\affiliation{Dipartimento di Fisica, Universit\`a di Genova, 16146 Genova, Italy}
	\affiliation{SPIN-CNR, 16146 Genova, Italy}
	\author{F. Cavaliere}
	\affiliation{Dipartimento di Fisica, Universit\`a di Genova, 16146 Genova, Italy}
	\affiliation{SPIN-CNR, 16146 Genova, Italy}
	\date{\today}
	
	\begin{abstract}
		The properties of prototypical examples of one-dimensional fermionic systems undergoing a sudden quantum quench from a gapless state to a (partially) gapped state are analyzed. By means of a Generalized Gibbs Ensemble analysis or by numerical solutions in the interacting cases, we observe an anomalous, non-monotonic response of steady state correlation functions as a function of the strength of the mechanism opening the gap. In order to interpret this result, we calculate the full dynamical evolution of these correlation functions, which shows a freezing of the propagation of the quench information (light cone) for large quenches. We argue that this freezing is responsible for the non-monotonous behaviour of observables. In continuum non-interacting models, this freezing can be traced back to a Klein-Gordon equation in the presence of a source term. We conclude by arguing  in favour of the robustness of the phenomenon in the cases of non-sudden quenches and higher dimensionality.
	\end{abstract}
	
	\maketitle
	Non-equilibrium quantum physics is at the heart of most relevant applications of solid state physics, such as transistors and lasers~\cite{vignale,Grosso:2014,Haug:2008}. More fundamentally, one of the main difficulties in studying many-body non-equilibrium quantum physics is represented by the unavoidable interactions that any quantum system has with its surroundings. This coupling is difficult to control and causes an effectively non-unitary evolution even on short time scales~\cite{weiss}. The recent advent of cold atom physics~\cite{bloch} allowed not only to access quantum systems characterized by weak coupling to the environment, but also to engineer Hamiltonians which show non-ergodic behavior~\cite{cradle,Kinoshita:2006}: the so called integrable systems~\cite{integrable}. Moreover, in the context of cold atom physics, it is possible to manipulate the parameters of the Hamiltonian in a time dependent and controllable fashion~\cite{Kinoshita:2006,Trotzky:2012,Cheneau:2012,Langen:2015,polkovnikov}. The combination of these three ingredients gave rise to a renewed interest in the physics of quantum quenches~\cite{Calabrese:2006,quench1,DAlessio:2016,Essler:2016,quench2}, which led to the birth of a new thermodynamic ensemble, the Generalized Gibbs Ensemble (GGE)~\cite{gge,gge1,gge2,gge3,Vidmar:2016,ggexp}. Quantum quenches have been studied in a wide range of systems with the property that a change in a parameter of the Hamiltonian deeply affects the physical properties of the system itself. Interaction quenches in Luttinger liquids~\cite{l1,l2,l3,l4,l5,l6,l7,l8,l9,l10,l11,l12,l13} and magnetic field quenches in the one-dimensional (1D) Ising model~\cite{i1,i2,i3,i4,i5,i6,i7,i8,i9,i10,i11} are prominent examples in this direction. Furthermore, at the level of free fermions, quantum quenches between gapped phases characterized by different Chern numbers have also been studied~\cite{c1,c2,c3,c4}. However, not much attention has been devoted to the study of quantum quenches between gapless and gapped states. A notable exception is represented by quantum quenches from a Luttinger liquid to a sine-Gordon model~\cite{sg1,sg2,sg3,sg4,sg5,sg6,sg7,sg8,sg9,sg10} and quantum time mirrors~\cite{tm}. However, the characterization of the main features of gapless-to-gapped quantum quenches is still lacking.
	
	In this Rapid Communication we consider paradigmatic examples of gapless 1D systems which get partially or completely gapped by a change in the parameters of the Hamiltonian. Namely, a spin-orbit coupled (SOC) quantum wire in the presence of an applied magnetic field~\cite{streda,seb,socold,pic} and a chain of spin-less 1D fermions. For the latter, the gapping quench mechanism is either induced by a staggered potential (SP) or by the sudden switch-on of fermion-fermion interactions~\cite{pd}.
	
	When the quench does not involve interactions, we consider both the lattice models and their continuum counterparts, describing the low-energy sector of a wide class of 1D systems. In these cases, the Hamiltonian $H$ can be written as $H(t)=
	\sum_k \Psi^\dag_k [\mathcal{H}_k+\theta(t)\Delta\sigma^x]\Psi_k$, where $\mathcal{H}_k$ is a family of $2\times2$ matrices indexed by the (quasi-) momentum $k$, characterized by a gapless spectrum. Here, $\sigma^x$ is the first Pauli matrix in the usual representation, $\Delta$ is the strength of the gap opening mechanism and $\theta(t)$ is the Heaviside function. Finally, $\Psi^\dag_k=(d^\dag_{a,k},d^\dag_{b,k})$ is a two-component momentum resolved Fermi spinor.
	
	In the case of the SOC wire, the indexes $a,b$ represent the spin projection along the quantization axis and $\Delta $ is proportional to the applied magnetic field. In the case of the SP model, the former labels left-/right- movers while $\Delta$ is proportional to the strength of the staggered potential. For the non-interacting cases we demonstrate that the quantity $M=\sum_k\langle\Psi^\dag_k\sigma^x\Psi_k\rangle_{GGE}/n$, where $\langle \cdot\rangle_{GGE}$ denotes the average on the associated GGE and $ n $ is the total number of particles in the system, exhibits, a maximum for a finite value of $\Delta$ and tends to the gapless value for $\Delta\rightarrow\infty$, meaning that the observable does not feel the quench for strong quenches. The same behavior also characterizes the scenario of gap opened by fermion-fermion interactions, both in the integrable and in the non-integrable case. In the non-integrable case, however, the results should be intended as valid in a long lived pre-thermal state. In order to interpret the result, we study, in the continuous non-interacting models, the time dependence of the correlation function $\mathcal{G}(x,t)=\langle\Psi^\dag(x,t)\sigma^x\Psi(0,t)\rangle$, where $\Psi^\dag(x,t)$ is the Fermi spinor, the average is performed with respect to the pre-quench ground state and $\lim_{t\rightarrow\infty}\mathcal{G}(0,t)=M$. For small quenches $\mathcal{G}(x,t)$ shows the propagation of a light cone conveying the information of the quench through the system while for large quenches the light cone {\em freezes}. In fact, $\mathcal{G}(x,t)$ is governed by a Klein-Gordon (KG) equation with a mass term $\propto\Delta^2$ and a source term $\propto\Delta$: the source term is responsible for finite values of $M$, while the mass term acts as a stiffness which hinders the generation and propagation of the cone. The same behaviour occurs also when interactions are quenched, providing strong evidence that the freezing of the light cone is responsible of the non-monotonous behaviour of observables. Finally, we conclude by analyzing the generality of the results.
	
	We begin analyzing the four non-interacting models, indexed by $i=1,..,4$. For the SOC quantum wire on the lattice we have $\mathcal{H}_k^{(1)}=2[1-\cos (k)]\sigma^0+\alpha\sin(k)\sigma^z$ (with $\sigma^0=I_{2\times2}$) and the gap opening time-dependent mechanism is given by the magnetic field $\Delta^{(1)}= B$. Here, the lattice constant has been set to 1 and $\alpha$ represents the spin-orbit coupling. The corresponding low-energy continuous theory is obtained by replacing $\mathcal{H}_k^{(1)}$ with $\mathcal{H}_k^{(2)}=k^2\sigma^0+\alpha k\sigma^z$, with gap opening parameter $\Delta^{(2)}=B$. We also discuss the SP model, with $\mathcal{H}_k^{(3)}=-2J\cos(k)\sigma^z$ and $\Delta^{(3)}$ the strength of the staggered potential. In this case, the sum is restricted to positive $k$ only. To obtain a low-energy theory for the SP model we expand around $k=\pi/2$, obtaining a Dirac cone with velocity $2J$ with a gap opening term $\Delta^{(4)}=\Delta^{(3)}$. The Hamiltonian density is $\mathcal{H}_k^{(4)}=-2J k\sigma^z$.
	
	We assume that, before the quench, the chemical potential is set to zero and the system is in its zero-temperature equilibrium ground state. We define $|\Phi^{(i)}_0(0)\rangle$ as the $i-$system ground state at $t=0$ of the corresponding pre-quench Hamiltonian. We introduce the unitary transformation $U^{(i)}_{0,k}$ satisfying $U^{(i)}_{0,k}\mathcal{H}_k^{(i)}U^{(i)\dag}_{0,k}=\mathrm{diag}\Big\{\epsilon_{+,0,k}^{(i)},\epsilon_{-,0,k}^{(i)}\Big\}$, with $\epsilon_{-,0,k}^{(i)}\leq\epsilon_{+,0,k}^{(i)}\ \forall k$, to get
	\begin{equation}
	|\Phi^{(i)}_0(0)\rangle=\prod_{k_1^{(i)}}^{k_2^{(i)}}\left(U^{(i)\dag}_{0,k}\Psi^{(i)\dagger}_k\right)_2|0^{(i)}\rangle.
	\end{equation}
	Here, $|0^{(i)}\rangle$ is the vacuum of the $i-$th Hamiltonian, $ k_{1/2}^{(i)}$ are fixed by the condition that only states with negative and zero energy are occupied, and the subscript $ 2 $ means that the second component of the spinor has to be considered. Note that the choice of the occupation of the zero energy modes is of no importance for the following since all results will be evaluated in the thermodynamic limit. Although $k_{1/2}^{(i)}$ are computed exactly in the calculations, here we only report the approximated relations $k_{1}^{(1)}\simeq k_{1}^{(2)}=-\alpha$, $k_{2}^{(1)}\simeq k_{2}^{(2)}=\alpha$ and $ k_1^{(3)}=k_1^{(4)}=0,\ k_2^{(3)}=k_2^{(4)}=\pi$~\cite{footnote:continuumSSH}.
	
	In order to get the time evolution of the system for $ t>0 $ we introduce a second unitary operator $U^{(i)}_{1,k}$ related to the post-quench Hamiltonian by $U^{(i)}_{1,k}[\mathcal{H}_k^{(i)}+\Delta^{(i)}\sigma^x]U^{(i)\dag}_{1,k}=\mathrm{diag}\{\epsilon_{+,1,k}^{(i)},\epsilon_{-,1,k}^{(i)}\}$, with $\epsilon_{-,1,k}^{(i)}\leq\epsilon_{+,1,k}^{(i)}\ \forall k$. In the Heisenberg representation, the time evolution of the systems is thus encoded in the Fermi spinor,
	\begin{equation}
	\Psi^{(i)}_k(t)=U^{(i)\dag}_{1,k}\mathrm{diag}\left\{e^{-i\epsilon_{+,1,k}^{(i)}t},e^{-i\epsilon_{-,1,k}^{(i)}t}\right\}U^{(i)}_{1,k}\Psi^{(i)}_k(0).\label{eq:evopsi}
	\end{equation}
	Long after the quench, each of the four systems considered reaches a steady state which is locally described by a GGE~\cite{gge}. The latter is constructed by considering as conserved quantities the occupation numbers $n^{(i)}_{k,j=1,2}$ of the energy levels of the corresponding post-quench Hamiltonian,
	\begin{equation}
	n^{(i)}_{k,j=1,2}=\left(\Psi^{(i)\dag}_kU^{(i)\dag}_{1,k}\right)_j\left(U^{(i)}_{1,k}\Psi^{(i)}_k\right)_j.
	\end{equation}
	Here, the subscript $j$ on the right-hand side means that the $j$-th component of the spinor must be considered. The GGE density matrices are hence given by
	\begin{equation}
	\rho^{(i)}=\frac{e^{-\sum_{k,j}\lambda_{k,j}n^{(i)}_{k,j}}}{\mathrm{Tr}\left\{e^{-\sum_{k,j}\lambda_{k,j}n^{(i)}_{k,j}}\right\}}.
	\end{equation}
	The Lagrange multipliers $\lambda_{k,j}$ are fixed by the condition $\langle \Phi_0^{(i)}(0) | n^{(i)}_{k,j}|\Phi_0^{(i)}(0)\rangle=\mathrm{Tr}\left\{ n^{(i)}_{k,j}\rho^{(i)}\right\}$.
	
	We can now compute the observables of interest. We first focus on $M^{(i)}=\sum_k\langle \Psi^{(i)\dag}_k\sigma^x \Psi^{(i)}_k\rangle_{GGE}/n^{(i)}$, with $ n^{(i)} $ the total number of particles in the $ i-$th system. All quantities can be evaluated analytically: the resulting expressions can be found in the Supplemental Material.
	
	\begin{figure}[htbp]
		\begin{center}

			\includegraphics[width=8.5cm,keepaspectratio]{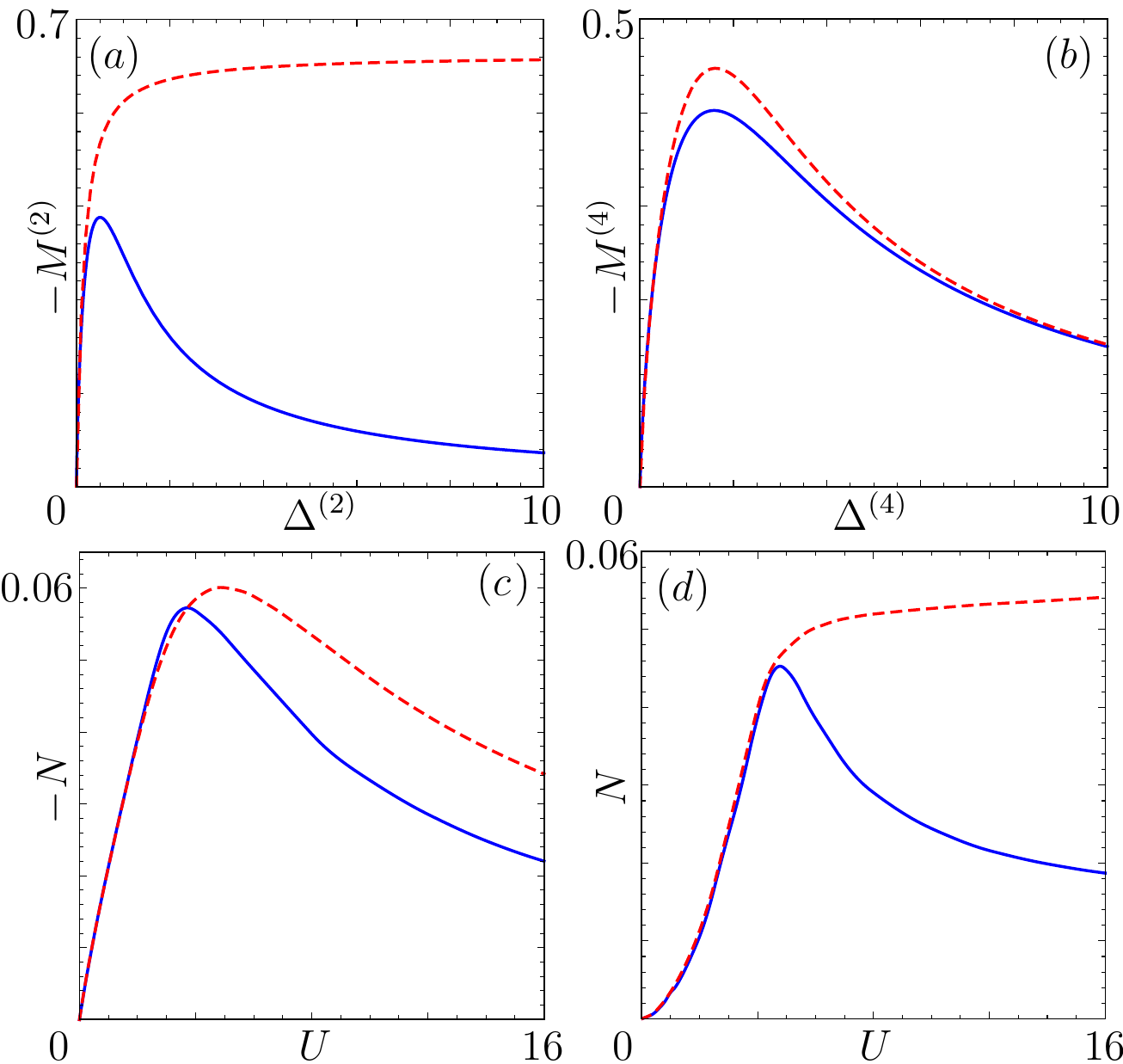}
			\caption{$(a)$ $M^{(2)}$ for the continuum SOC wire as a function of $\Delta^{(2)}$ with $ \alpha=1 $; $(b)$ $M^{(4)}$ for the continuum SP model as a function of $\Delta^{(4)}$ and $J=1$; $(c)$ $N$ for an interaction-quenched fermion chain, as a function of $U$ with $V=0$ and $J=1$; $(d)$ $N$ for an interaction-quenched fermion chain, as a function of $U$ with $V=U$ and $J=1$. In all panels, the solid lines represent the quenched long-time limit, the dashed lines show the results evaluated with an effective thermal model (see text).}
			\label{fig:1}
		\end{center}
	\end{figure}
	
	The results for the continuum SOC and SP models are shown by solid lines in Fig.~\ref{fig:1}, panels (a) and (b), as $\Delta^{(i)}$ is increased. In both cases, $M^{(i)}$ is non-monotonous, increasing up to a maximum before dropping to the pre-quench value. The results for the two lattice models are qualitatively similar and are reported in the Supplemental Material. A first interpretation of the phenomenon is the following: For infinitesimal $\Delta^{(i)}$ we do not expect any difference between a sudden quench and an adiabatic switching on of the gap opening mechanism. Thus, the systems begins to magnetize. On the other hand, when $\Delta^{(i)}$ strongly exceeds the kinetic energy, $M^{(i)}$ is conserved and hence it remains at its pre-quench value. A maximum for finite $\Delta^{(i)}$ is thus expected.
	
	We compare the GGE results with those obtained by an effective thermal ensemble at a given effective temperature $\beta_{\mathrm{eff}}^{-1}(\Delta^{(i)})$. The latter is obtained by solving
	\begin{equation}
	\langle\Phi^{(i)}_0(0)|H^{(i)}_{pq}|\Phi^{(i)}_0(0)\rangle=\frac{\mathrm{Tr}\left\{e^{-\beta_{\mathrm{eff}}(\Delta^{(i)})(H^{(i)}_{pq}-\mu(\Delta^{(i)})n^{(i)})}H^{(i)}_{pq}\right\}}{\mathrm{Tr}\left\{e^{-\beta_{\mathrm{eff}}(\Delta^{(i)})(H^{(i)}_{pq}-\mu(\Delta^{(i)})n^{(i)})}\right\}}\label{efft}
	\end{equation}
	for the effective temperature. Here, $H^{(i)}_{pq}=H^{(i)}(t>0)$ is the post-quench Hamiltonian and $\mu(\Delta^{(i)})$ is the Lagrange multiplier ensuring particle number conservation. The effective-temperature magnetization is shown in Figs.~\ref{fig:1}(a,b) as a dashed line. While for the SP model there is a good qualitative agreement, for the SOC wire the disagreement is dramatic as the effective temperature magnetization saturates to a non-zero value. The mechanism behind the non-monotonous behaviour of $M^{(i)}$ is hence not effective heating. 
	
	To get a deeper understanding, we now focus on the continuum models and introduce the Green's function
	\begin{equation}
	\label{eq:Gf}
	\mathcal{G}^{(i)}(x,t)=\langle\Phi^{(i)}_0(0)|\Psi^{(i)\dagger}(x,t)\sigma^{x}\Psi^{(i)}(0,t)|\Phi^{(i)}_0(0)\rangle\, .
	\end{equation}
	Clearly, $\mathcal{G}^{(i)}(0,t)=M^{(i)}(t)$. $\mathcal{G}^{(i)}(x,t)$ satisfies, for $t>0$, an inhomogeneous KG equation
	\begin{equation}
	\label{eq:KG}
	\left(\partial_x^2-\frac{1}{4u_i^2}\partial_t^2\right)\mathcal{G}^{(i)}(x,t)=\lambda_i^2 \mathcal{G}^{(i)}(x,t)+\lambda_i \phi_i(x)\, ,	
	\end{equation}
	where $\lambda_i=\Delta^{(i)}/u_i$ (with $u_2=\alpha$, $u_4=J$) and the source term is $\phi_i(x)=i\partial_x\langle\Psi^{(i)\dagger}(x,0)\mathcal{M}^{(i)}\Psi^{(i)}(0,0)\rangle_0$, with $ \mathcal{M}^{(2)}=\sigma^{z} $ and $ \mathcal{M}^{(4)}=\sigma^{y}$. Equation~\eqref{eq:KG} is solved with the pre-quench boundary-value condition $\mathcal{G}^{(i)}(x,0)=0$. 
	\begin{figure}[htbp]
		\begin{center}
			\includegraphics[width=8.5cm,keepaspectratio]{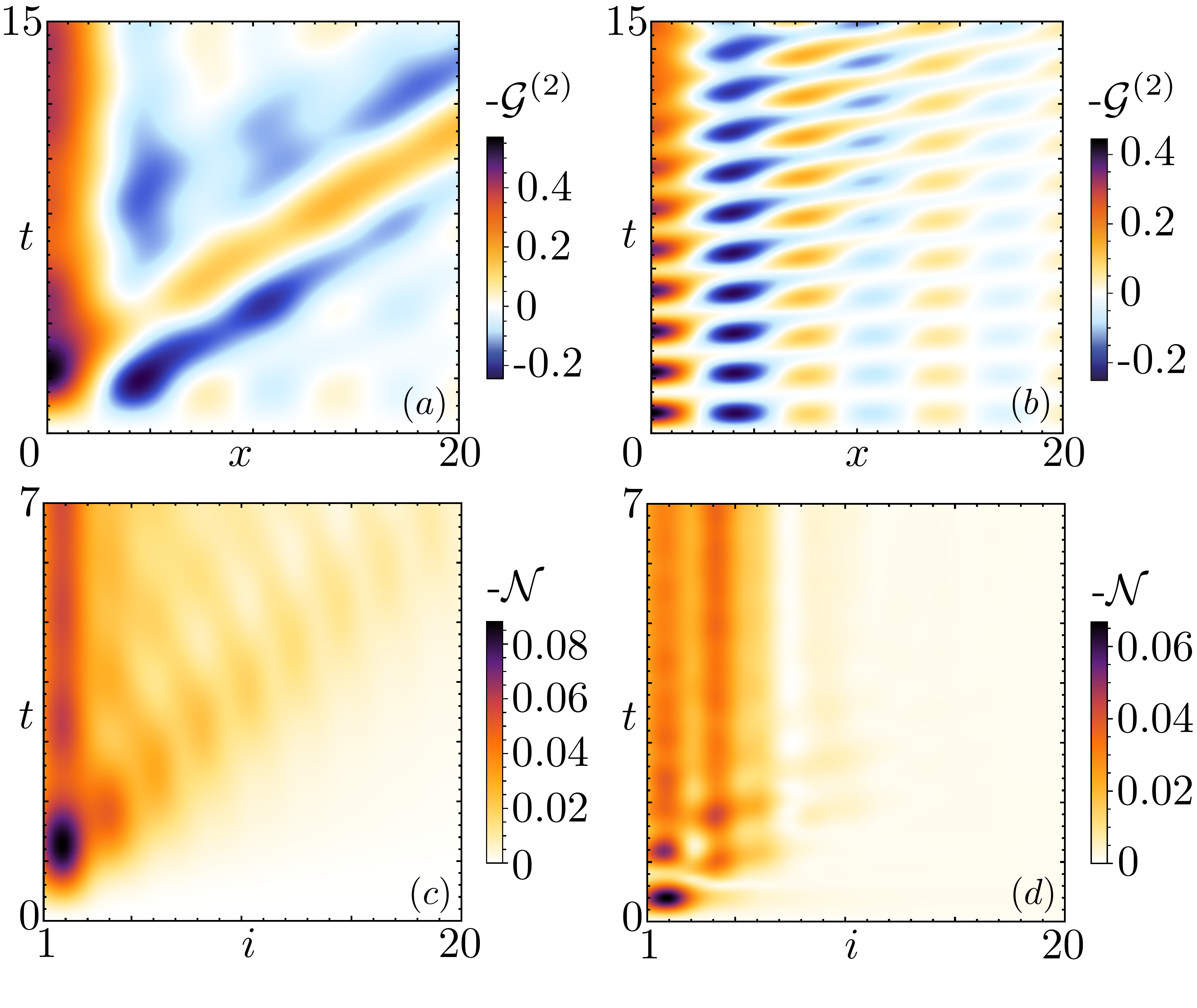}
			\caption{(a) Density plot of $-\mathcal{G}^{(2)}(x,t)$ as a function of $x$ and $t$ for $\Delta^{(2)}=0.3$ and $\alpha=1$; (b) same as in (a) but for $\Delta^{(2)}=2$ and $\alpha=1$; (c) Density plot of $-\mathcal{N}(i,t)$ as a function of $i$ and $t$ for $U=2$, $V=0$ and $J=1$; (d) same as in (c) but for $U=16$, $V=0$ and $J=1$.}
			\label{fig:2}
		\end{center}
	\end{figure}
	Equation~\eqref{eq:KG} supports a steady-state solution for $t\to\infty$ and it can be checked (see Supplemental Material) that $\lim_{t\to\infty}\mathcal{G}^{(i)}(0,t)=M^{(i)}$. Therefore, analyzing the time evolution of $\mathcal{G}^{(i)}(x,t)$ can shed light on the dynamics leading to the non-monotonous magnetization. This is shown in Fig.~\ref{fig:2} for the SOC wire model, in the case of (a) small quench $\Delta^{(2)} =0.3$ and (b) large quench $\Delta^{(2)}=2$. For a small quench, $ \mathcal{G}^{(2)}(x,t) $ exhibits a typical light-cone behavior~\cite{Calabrese:2006,l1,l2} and information of the quench is therefore able to propagate throughout the system.  This leaves a finite ``trail" in $x=0$, which eventually results in a finite value of $M^{(2)}$ at large times. On the other hand, the response of $ \mathcal{G}^{(2)}(x,t) $ to the ``shock" induced by a large quench is dramatically different. Indeed, in this regime, $ \mathcal{G}^{(2)}(x,t) $ is characterized by weakly damped and almost stationary oscillations both in space and in time, which strongly hinder the propagation of the information through the system and leads to both a slowdown and an overall suppression of the light cone. The Green's function $ \mathcal{G}^{(2)}(x,t) $ oscillates around its pre-quench initial one and reaches in the long-time limit a value very close to the latter. This phenomenon can thus be interpreted as an effective freezing of the light cone. The same qualitative behavior is observed also for the continuum SP model, not shown here. We thus attribute the emergence of the non-monotonous behavior of $M^{(i)}$ as a function of $\Delta^{(i)}$ to the competition between the propagating and freezing regimes. The identification of a freezing regime for large quenches constitutes a crucial result and - as we shall show in the last part of the paper - is a quite universal and robust feature of gap-opening sudden quench models. It represents a new concept in the physics of quantum quenches: Even though the gap is not able to dynamically introduce a length scale in the correlation functions, it dramatically influences the light-cone propagation.
	

	In order to get a picture of the effect, it is worth to notice that, in a simple mechanical interpretation~\cite{Courant:1953}, the KG equation represents the transverse vibrations of a string driven by a force $\propto\lambda_i$, embedded into an elastic medium of elastic constant $\propto\lambda_i^2$. When the medium is slack, vibrations can propagate almost without disturbance, while in a stiff medium the wave propagation is strongly suppressed. The turning point turns out to be for $\lambda_i\sim 1$, which corresponds to the location of the maximum of $M^{(i)}$ shown in Fig.~\ref{fig:1}($a$, $b$). Therefore, when the gap becomes comparable to the average kinetic energy scale, the freezing of the light cone begins to occur. Thanks to Wick's theorem, a similar behaviour characterizes all higher order correlators. This issue is relevant, since some of those correlators are either easier to numerically evaluate in the interacting systems we will analyze, or experimentally more accessible.
	
	We now turn to the lattice model described by $\mathcal{H}_k^{(3)}$ where, instead of switching on a staggered potential, a sudden quench of the nearest-neighbour interaction $U$ and/or of the next-to-nearest neighbour interaction $V$ is performed. For $U,V\gtrsim J$ interactions can open a gap in the spectrum. In addition, when $V>0$ the model is non-integrable. We turn to a numerical evaluation employing the DMRG technique.~\cite{dmrg1,dmrg2,dmrg3,dmrg4}.
	
	Since the model is invariant under rotations in the spinor space, we analyze the long-time (stationary) limit $N$ of the correlation function $\mathcal{N}(1,t)$, defined by $\mathcal{N}(i,t)=\langle(n_0(t)-1/2)(n_{i}(t)-1/2)\rangle_0-\langle(n_0(0)-1/2)(n_{i}(0)-1/2)\rangle_0$, where $n_i(t)$ is the (time-resolved) occupation number of the $i$-th site and $ \langle \cdot \rangle_0 $ represents the average with respect to the pre-quench ground state. Results are shown in Fig.~\ref{fig:1}(c) for the integrable case $V=0$ (solid line). $N$ follows the same qualitative behaviour of the magnetization in the non-interacting models, rising for small quenches up to a maximum value. As the gap size increases over the crossover point, $N$ begins to decrease and tends (not shown) to the pre-quench value for very large $U$. Even when integrability is lost, as is the case of Fig.~\ref{fig:1}(d) ($V=U$, solid line), the qualitative picture remains, on the accessible time-scales, the same. In both cases, a description in terms of an effective temperature (dashed lines) fails to reproduce the results.
	
	Also in this model a competition between a propagation and a freezing regime for the light cone occurs. To show this fact, we consider the correlation function $\mathcal{N}(i,t)$, shown in Figs.~\ref{fig:2}(c,d). For small quenches with $U\le 2J$ one clearly observes a propagation of the quench information spreading through the system. On the other hand, large quenches with $U>2J$ display a sharp freezing of the light cone. Thus, the freezing of the light cone is a generic feature of systems subject to quenches opening large gaps in the spectrum.
	
	To further support the idea that this mechanism is robust and represents a generic feature, we have checked that the results obtained here are valid even in higher dimensions. We have considered the paradigmatic case of a quench of a magnetic field applied to a Rashba-coupled two-dimensional electron gas~\cite{Bercioux:2015}. Also in this case, the long-time magnetization shows a non-monotonous behaviour as a function of the magnetic field, increasing to a maximum before eventually turning to the pre-quench value for large quenches - see Supplemental Material. The results are robust even with respect to the rapidity of the quench. We have studied the continuum SOC wire model when the magnetic field linearly ramps from 0 - see Supplemental Material for details. For longer ramps, the asymptotic value of $M^{(2)}$ for large $\Delta^{(2)}$ increases, but the non-monotonous behaviour of the magnetization persists.
	
	In conclusion, a non-monotonic behavior of observables characterizes a wide range of gapless-to-(partially) gapped quantum quenches, both for sudden and non-sudden protocols, integrable and non-integrable models and not only in one spatial dimension. It is the hallmark of a peculiar phenomenon, namely the freezing of the light cone which conveys the quench information through the system. This freezing results in a state described by a GGE which differs from effective thermal states, in some cases even dramatically, thus providing an experimentally accessible way to test the GGE physics. In non-interacting models, the freezing of the light cone is captured by a KG equation, which provides an intuitive interpretation of the behaviour of the system in terms of a simple continuum mechanical model. As a limit for the universality of the physics described, we point out that we do not expect to observe the effects when the gap is opened by merging of crossings, as relevant, for example, for Weyl semimetals~\cite{weyl}, or for the models discussed in Refs.~\cite{herbut,Montambaux:2009}. A static fermion-fermion interaction, which could be taken into account by means of bosonization~\cite{g1,voit,Gambetta:2015,traverso} or DMRG~\cite{dmrg1,dmrg2} for instance, is expected to renormalize the gap to larger values~\cite{meng}, so we expect the phenomenon to persist with a shifted and renormalized maximum~\cite{giamarchi}.
	
	\begin{acknowledgments}
		The authors would like to thank Markus Heyl for useful discussions. N. T. Z. gratefully acknowledges financial support by the DFG (Grants No. SPP1666 and No. SFB1170 “ToCoTronics”), the Helmholtz Foundation (VITI), and the ENB Graduate school on “Topological Insulators”. D. M. K. was supported by the Basic Energy Sciences Program of the U. S. Department of Energy under Grant No. SC-0012375 and by DFG KE 2115/1-1. Simulations for the interacting model were performed with
		computing resources granted by RWTH Aachen University under project rwth0013. 
	\end{acknowledgments}

\pagebreak
\widetext
\begin{center}
	\textbf{\large Supplemental Material for “Non-monotonic response and light-cone freezing in gapless-to-(partially) gapped quantum quenches of fermionic systems”}
\end{center}

\setcounter{equation}{0}
\setcounter{figure}{0}
\setcounter{table}{0}
\makeatletter
\renewcommand{\theequation}{S\arabic{equation}}
\renewcommand{\thefigure}{S\arabic{figure}}

\renewcommand{\bibnumfmt}[1]{[S#1]}
\renewcommand{\citenumfont}[1]{S#1}

\section{Steady state magnetization}
\subsection{Diagonalization of a generic $ 2\times2 $ Hermitian matrix}\label{SM:subsec:general_diag}
In order to set the conventions, we begin this Section by briefly summarizing the diagonalization procedure for a generic $ 2\times2 $ Hermitian matrix, 
\begin{equation}
\mathcal{H}=\begin{bmatrix}
h_{11} &h_{12}\\
h_{12}^* &h_{22}
\end{bmatrix},\label{SM:eq:generalH}
\end{equation}
with $ h_{11},\,h_{12}\in\mathbb{R} $ and $ h_{12}\in\mathbb{C} $. We first focus on the case $ h_{12}\neq0 $. Then, the eigenvalues of $ \mathcal{H} $ are
\begin{equation}
\epsilon_{\pm}=\frac{1}{2}\left(h_{11}+h_{22}\right)\pm D,
\end{equation}
where $ D=\sqrt{(h_{11}-h_{22})^2+4|h_{12}|^2}/2 $. The Hamiltonian of Eq.~\eqref{SM:eq:generalH} can be diagonalized by means of the unitary matrix $ U $,
\begin{equation}
UHU^\dagger=\begin{bmatrix}
\epsilon_+ &0\\
0 &\epsilon_-
\end{bmatrix},\qquad\text{with}\qquad U=\begin{bmatrix}
A_- &-A_-\frac{\epsilon_--h_{22}}{h_{12}^*}\\
-A_+\frac{\epsilon_+-h_{11}}{h_{12}} &A_+
\end{bmatrix}\quad\text{and}\quad\varepsilon_+>\varepsilon_-,
\label{SM:eq:U}
\end{equation}
where we have introduced the coefficients 
\begin{equation}
A_+=\frac{|h_{12}|}{\sqrt{(\epsilon_+-h_{11})^2+|h_{12}|^2}}\qquad\text{and}\qquad A_-=\frac{|h_{12}|}{\sqrt{(\epsilon_--h_{22})^2+|h_{12}|^2}}.
\end{equation}
On the other hand,  in the case $ h_{12}=0 $, the unitary matrix $ U $ that transforms $ \mathcal{H} $ in the diagonal form of Eq.~\eqref{SM:eq:U}, i.e. with $ \epsilon_+>\epsilon_- $, is
\begin{equation}
\label{SM:eq:Udiag}
U=
\begin{cases}
I_{2\times2}\theta(h_{11}-h_{22})+i\sigma^y\theta(h_{22}-h_{11}), & \text{if }h_{11}\neq h_{22},\\
\frac{1}{\sqrt{2}}\left(I+i\sigma^{y}\right), &\text{if }h_{11}= h_{22},
\end{cases}
\end{equation}
with $ I_{2\times2} $ the $ 2\times2 $ identity matrix and $ \sigma^{y} $ the $ y $ Pauli matrix in the usual representation.
\subsection{Some general formulas on the calculation of $ M $ in 1D systems}\label{SM:sec:generalM}
As stated in the main text, the Hamiltonian of both the SOC wire and the SP model can be written as 
\begin{equation}
H^{(i)}(t)=\sum_k \Psi^{(i)\dagger}_k [\mathcal{H}^{(i)}_k+\theta(t)\Delta^{(i)}\sigma^x]\Psi^{(i)}_k.\label{SM:eq:H}
\end{equation}
Here, $ \Psi^{(i)\dagger}_k=\left(d_{a,k}^{(i)\dagger},d_{b,k}^{(i)\dagger}\right) $ is a two-component momentum resolved Fermi spinor. In the case of the SOC wire ($ i=\{1,2\} $), the indexes $ a $ and $ b $ represent the positive and negative spin projections along the quantization axis, respectively, while in the case of the SP model ($ i=\{3,4\} $) they are associated with the left-/right-movers. The pre-quench single-mode Hamiltonian $ \mathcal{H}^{(i)}_k $ can always be written in a diagonal form with eigenvalues $ \epsilon^{(i)}_{\pm,0,k} $ such that $ \epsilon^{(i)}_{-,0,k}\leq\epsilon^{(i)}_{+,0,k},\, \forall k $, by means of a unitary matrix [see Eqs.~\eqref{SM:eq:U} and~\eqref{SM:eq:Udiag}]. In particular, for all the cases considered in this paper the latter takes the form
\begin{equation}
U^{(i)}_{0,k}=\begin{bmatrix}
a^{(i)}_{0,k} &b^{(i)}_{0,k}\\
-b^{(i)*}_{0,k} &a^{(i)}_{0,k}
\end{bmatrix},\label{SM:eq:U0}
\end{equation}
where the coefficients $ a^{(i)}_{0,k}\in\mathbb{R} $ and $ b^{(i)}_{0,k}\in\mathbb{C} $ are determined by Eqs.~\eqref{SM:eq:U} and~\eqref{SM:eq:Udiag}. Moreover, $ U_{0,k}^{(i)}\mathcal{H}^{(i)}_k U_{0,k}^{(i)\dagger}=\text{diag}\Big\{\epsilon_{+,0,k}^{(i)},\epsilon_{-,0,k}^{(i)}\Big\}  $. For $ t<0 $ the diagonalized Hamiltonian reads
\begin{equation}
H^{(i)}(t<0)=\sum_k \left[\epsilon^{(i)}_{-,0,k}d^{(i)\dagger}_{v,0,k}d^{(i)}_{v,0,k}+\epsilon^{(i)}_{+,0,k}d^{(i)\dagger}_{c,0,k}d^{(i)}_{c,0,k}\right],
\end{equation}
where the conduction and valence band operators, $ d^{(i)}_{c,0,k} $ and $ d^{(i)}_{v,0,k} $, are defined by
\begin{equation}
\Phi_{0,k}^{(i)}=U^{(i)}_{0,k}\Psi^{(i)}_k=\begin{bmatrix}
d_{c,0,k}^{(i)} \\
d_{v,0,k}^{(i)}
\end{bmatrix}.
\end{equation}
In all cases considered we set the chemical potential to zero and assume the $ i- $th system to be in its pre-quench zero-temperature equilibrium ground state, $ |\Phi^{(i)}_{0}(0)\rangle $. Therefore, for $ t<0 $, the bands are filled up to the linear crossing and $|\Phi^{(i)}_{0}(0)\rangle $ is defined as
\begin{equation}
|\Phi^{(i)}_{0}\rangle=\prod_{k_1^{(i)}}^{k_2^{(i)}}\left(\Phi^{(i)\dagger}_{0,k}\right)_2|0^{(i)}\rangle=\prod_{k_1^{(i)}}^{k_2^{(i)}}\left(U^{(i)\dag}_{0,k}\Psi^{(i)\dagger}_k\right)_2|0^{(i)}\rangle,\label{SM:eq:GS0}
\end{equation}
with $ |0^{(i)}\rangle $ the vacuum of the $ i- $th system and  $ k_{1,2}^{(i)} $ determined by imposing that only negative energy states are filled. Here, the subscript $ 2 $ means that the second component of the spinor has to be considered.\\

We now turn to the regime with $ t>0 $. The post-quench single-mode Hamiltonian $ \mathcal{H}^{(i)}_k+\Delta^{(i)} \sigma^x $ is diagonalized by the unitary matrix
\begin{equation}
U^{(i)}_{1,k}=\begin{bmatrix}
a^{(i)}_{1,k} &b^{(i)}_{1,k}\\
-b^{(i)*}_{1,k} &a^{(i)}_{1,k}
\end{bmatrix},
\end{equation}
with $ a^{(i)}_{1,k}\in\mathbb{R} $ and $ b^{(i)}_{1,k}\in\mathbb{C} $ determined again by Eqs.~\eqref{SM:eq:U} and~\eqref{SM:eq:Udiag}, and  $ U_{1,k}^{(i)}[\mathcal{H}^{(i)}_k+\Delta^{(i)}\sigma^x] U_{1,k}^{(i)\dagger}=\text{diag}\{\epsilon_{+,1,k}^{(i)},\epsilon_{-,1,k}^{(i)}\}  $. The total Hamiltonian thus becomes
\begin{equation}
H^{(i)}(t>0)=\sum_k \left[\epsilon^{(i)}_{-,1,k}d^{(i)\dagger}_{v,1,k}d^{(i)}_{v,1,k}+\epsilon^{(i)}_{+,1,k}d^{(i)\dagger}_{c,1,k}d^{(i)}_{c,1,k}\right],
\end{equation}
where $ \epsilon^{(i)}_{-,1,k}\leq\epsilon^{(i)}_{+,1,k},\ \forall k $, with the new conduction and valence band fermionic operators, $  d^{(i)}_{c,1,k} $ and $  d^{(i)}_{v,1,k} $, given by
\begin{equation}
\Phi^{(i)}_{1,k}=U^{(i)}_{1,k}\Psi_k^{(i)}=\begin{bmatrix}
d^{(i)}_{c,1,k}\\
d^{(i)}_{v,1,k}
\end{bmatrix}.
\label{SM:eq:Phi1gen}
\end{equation}
We now evaluate the magnetization of system along the direction of the applied magnetic field in the SOC wire or the staggered magnetization in the SP model within the framework of the GGE~\cite{Vidmar:2016}. To do this, it is sufficient to know the average over the pre-quench ground state $ |\Phi_0^{(i)}\rangle $, denoted by  $ \langle\cdot\rangle_0 $, of the occupation numbers $ n^{(i)}_{k,j} $ of the energy levels of the corresponding post-quench Hamiltonian, given by
\begin{equation}
n^{(i)}_{k,j=1,2}=\left(\Psi^{(i)\dagger}_k U^{(i)\dagger}_{1,k}\right)_j \left( U^{(i)}_{1,k}\Psi^{(i)}_k\right)_j.
\end{equation}
Since all the $ n^{(i)}_{k,j} $ commute with the post-quench Hamiltonian, they are conserved for $ t>0 $ and, therefore, $ \langle n^{(i)}_{k,j}\rangle_{0}=\langle n^{(i)}_{k,j}\rangle_{GGE} $. 
We obtain
\begin{subequations}
	\label{SM:eq:nGGE}
	\begin{align}
	\langle n^{(i)}_{k,1}\rangle_0&=\langle n^{(i)}_{k,1}\rangle_{GGE}=\left|-a^{(i)}_{1,k}b^{(i)}_{0,k}+a^{(i)}_{0,k}b^{(i)}_{1,k}\right|^2\langle d^{(i)\dagger}_{v,0,k} d^{(i)}_{v,0,k}\rangle_0,\\
	\langle n^{(i)}_{k,2}\rangle_0&= \langle n^{(i)}_{k,2}\rangle_{GGE}=\left|a^{(i)}_{1,k}a^{(i)}_{0,k}+b^{(i)}_{0,k}b^{(i)*}_{1,k}\right|^2\langle d^{(i)\dagger}_{v,0,k} d^{(i)}_{v,0,k}\rangle_0,
	\end{align}
\end{subequations}
where the averages $ \langle d^{(i)\dagger}_{v,0,k} d^{(i)}_{v,0,k}\rangle_0 $ can be easily evaluated from Eq.~\eqref{SM:eq:GS0}.
Using Eq.~\eqref{SM:eq:Phi1gen} and the fact that $ \langle d^{(i)\dagger}_{c,1,k}d^{(i)}_{v,1,k}\rangle_{GGE}=\langle d^{(i)\dagger}_{v,1,k}d^{(i)}_{c,1,k}\rangle_{GGE}=0 $, one gets the steady state magnetization (SOC wire) or the staggered magnetization (SP)
\begin{align}
M^{(i)}=\frac{1}{n^{(i)}}\sum_k \langle\Psi^{(i)\dagger}_k\sigma^x\Psi^{(i)}_k\rangle_{GGE}=\frac{1}{\left|k^{(i)}_1\right|+\left|k^{(i)}_2\right|}\int_{k^{(i)}_{1}}^{k^{(i)}_2}dk\,\left(a^{(i)}_{1,k}b^{(i)}_{1,k}+a^{(i)}_{1,k}b^{(i)*}_{1,k}\right)\left(\langle n^{(i)}_{k,1}\rangle_{GGE}-\langle n^{(i)}_{k,2}\rangle_{GGE}\right),\label{SM:eq:M}
\end{align}
where $ n^{(i)}=L(|k^{(i)}_1|+|k^{(i)}_2|)/(2\pi) $ is the total number of particles in the $ i- $th system and $ L $ is its length. Furthermore, in the last step, the thermodynamic limit has been performed.
\subsection{Spin-orbit coupled quantum wire}
We now explicitly apply the general discussion of Sec.~\ref{SM:sec:generalM} to the quench of the external magnetic field in a SOC wire. We start by introducing the real space lattice Hamiltonian, imposing periodic boundary conditions:
\begin{equation}
H=\sum_{j=1}^L \left\{ \Psi^\dag_j \left[2I_{2\times2}+B\theta(t)\sigma^x\right]\Psi_j + \Psi^\dag_j\left[\frac{i\alpha}{2}\sigma^z-I_{2\times2}\right]\Psi_{j+1}+h.c.\right\},
\end{equation} 
where $ L $ is the total length of the system and we set the lattice spacing to 1. In this case the pre-quench single-mode Hamiltonians $ \mathcal{H}^{(i)}_k $, with $ i=\{1,2\} $, are
\begin{subequations}
	\begin{align}
	\mathcal{H}_k^{(1)}&=2[1-\cos (k)]I_{2\times2}+\alpha\sin(k)\sigma^z,\qquad\text{with } k\in[-\pi,\pi),\\
	\mathcal{H}^{(2)}_k&=k^2I_{2\times2}+\alpha k \sigma^z,
	\end{align}
\end{subequations}
for the lattice and the low-energy continuous models, respectively. Note that, because of the translational invariance, $ \Psi_j = \sum_k e^{ikj}\ \Psi_k/\sqrt{L} $. Although the pre-quench single-mode Hamiltonian $ \mathcal{H}^{(i)}_k $ is already diagonal, it can be more conveniently rewritten in terms of conduction and valence band Fermi operators using Eq.~\eqref{SM:eq:U0}. In this case the coefficients of the unitary matrix $ U^{(i)}_{0,k} $ are 
\begin{equation}
a^{(i)}_{0,k}=[1-\delta_{k,0}]\theta(k)+\frac{\delta_{k,0}}{\sqrt{2}} \qquad b^{(i)}_{0,k}=[1-\delta_{k,0}]\theta(-k)+\frac{\delta_{k,0}}{\sqrt{2}},	\label{SM:eq:ab0SOC}
\end{equation}
while the conduction and valence energy bands are
\begin{subequations}
	\begin{align}
	\epsilon_{\pm,0,k}^{(1)}&=2[1-\cos(k)]\pm\alpha|\sin(k)|,\\
	\epsilon_{\pm,0,k}^{(2)}&=k^2\pm\alpha|k|.
	\end{align}
\end{subequations}
From $ \epsilon^{(i)}_{-,0,k}=0 $ it follows that $ k_{1/2}^{(1)}=\mp2\arctan[\alpha/2] $ and $ k_{1/2}^{(2)}=\mp\alpha $.  \\

In the post-quench regime $ t>0 $, the unitary matrix $ U^{(i)}_{1,k} $ which diagonalizes the single-mode Hamitonian $ \mathcal{H}^{(1)}_k+B\sigma^x $ has coefficients
\begin{subequations}
	\label{SM:eq:ab}
	\begin{align}
	a^{(1)}_{1,k}&=\frac{B}{\sqrt{\left[D^{(1)}_k-\alpha\sin(k)\right]^2+B^2}},\qquad &&b^{(1)}_{1,k}=\frac{D^{(1)}_k-\alpha \sin(k)}{\sqrt{\left[D^{(1)}_k-\alpha\sin(k)\right]^2+B^2}},\label{SM:eq:abL}\\
	a^{(2)}_{1,k}&=\frac{B}{\sqrt{\left[D^{(2)}_k-\alpha k\right]^2+B^2}},\qquad &&b^{(2)}_{1,k}=\frac{D^{(2)}_k-\alpha k}{\sqrt{\left[D^{(2)}_k-\alpha k\right]^2+B^2}},\label{SM:eq:abC}
	\end{align}
\end{subequations}
where $ D^{(1)}_k=\sqrt{\alpha^2 \sin^2(k)+B^2} $ and $ D^{(2)}_k=\sqrt{\alpha^2 k^2+B^2} $, while the post-quench conduction and valence bands are 
\begin{subequations}
	\begin{align}
	\epsilon_{\pm,1,k}^{(1)}&=2[1-\cos(k)]\pm D^{(1)}_k,\\
	\epsilon_{\pm,1,k}^{(2)}&=k^2\pm D^{(2)}_k.
	\end{align}
\end{subequations}

\begin{figure}[h]
	\centering
	\includegraphics[width=0.5\linewidth]{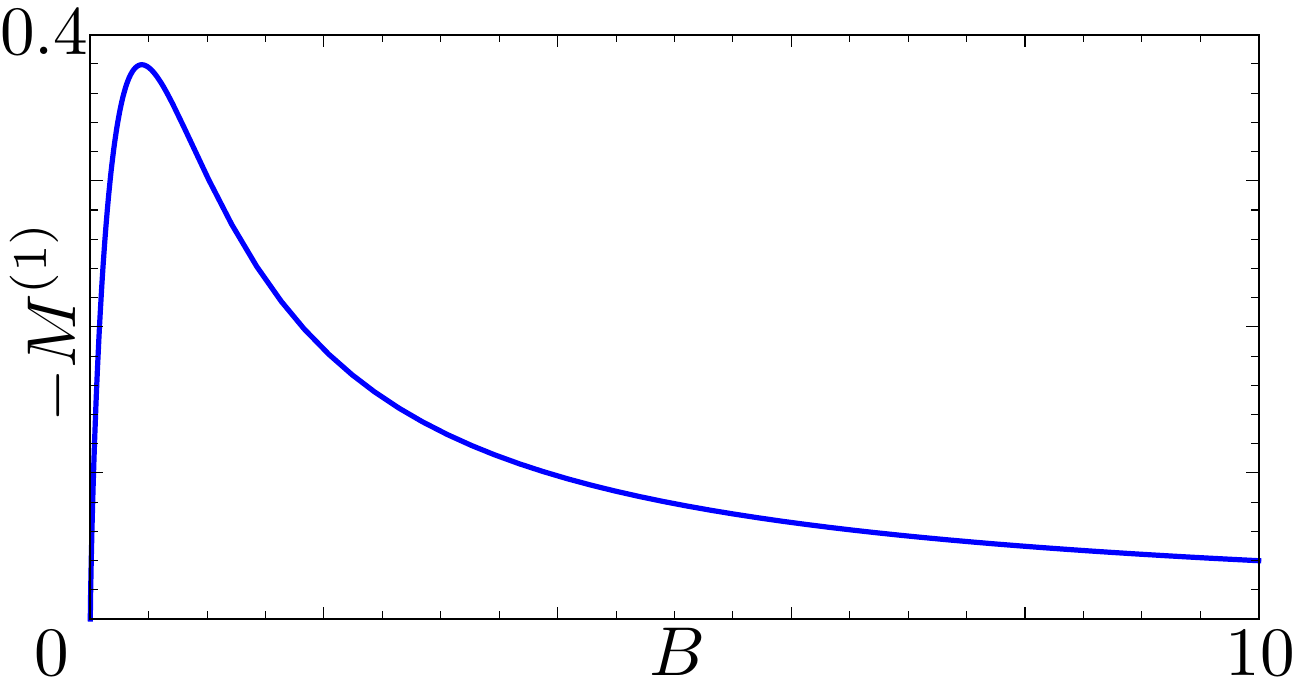}
	\caption{$ M^{(1)} $ for the lattice SOC wire as a function of $ \Delta^{(1)}=B $ and $ \alpha = 1 $.}
	\label{fig:socwirel}
\end{figure}

From Eq.~\eqref{SM:eq:M} one immediately obtains the steady state magnetization along the direction of the applied magnetic field in the thermodynamic limit,
\begin{equation}\label{SM:eq:MSOC}
M^{(i)}=\frac{1}{\left|k^{(1)}_1\right|+\left|k^{(1)}_2\right|}\left(\int_{k^{(i)}_1}^{0}-\int_{0}^{k^{(i)}_2}\right)dk\,2a^{(i)}_{1,k}b^{(i)}_{1,k} \left[\left(a^{(i)}_{1,k}\right)^2-\left(b^{(i)}_{1,k}\right)^2\right].
\end{equation}
For the lattice model, using Eq.~\eqref{SM:eq:abL}, we have
\begin{equation}
M^{(1)}
=\frac{1}{\left|k^{(1)}_1\right|+\left|k^{(1)}_2\right|}\frac{B}{2\sqrt{\alpha^2+B^2}}\log\left(\frac{Z_+}{Z_-}\right),
\end{equation}
with
\begin{equation}
Z_\pm=(\sqrt{\alpha^2+B^2}\mp\alpha)^2\left[\sqrt{\alpha^2+B^2}\pm\alpha\frac{4-\alpha^2}{4+\alpha^2}\right]^2.
\end{equation}
The behavior of $ M^{(1)} $ is shown in Fig~\ref{fig:socwirel}. On the other hand, for the low-energy continuous model one gets from Eq.~\eqref{SM:eq:abC}
\begin{equation}
M^{(2)}
=-\frac{B}{4\alpha^2}\log\left[1+2\frac{\alpha^4}{B^2}+\frac{\alpha^8}{B^4} \right].
\end{equation}

\subsection{Staggered Potential model}
In this Section we apply the discussion of Sec.~\ref{SM:sec:generalM} to the quench of the staggered potential in the SP model. The real space lattice Hamiltonian reads
\begin{equation}
H=-J\sum_{j=1}^L \left[\left(c^\dag_jc_{j+1}+h.c.\right)+(-1)^j\delta(t)c^\dag_jc_{j}\right],\label{SM:eq:H_SP}
\end{equation}
where $ c_j $ are annihilation operators for spinless fermions on the site $ j $ of the lattice, $ L $ is the length of the system and we set the lattice spacing to 1. Introducing the spinor $ \Psi^\dag_k = \left(c^\dag_k,c^\dag_{k-\pi}\right) $, with $ c_j=\sum_k e^{ikj} c_k/\sqrt{L} $, we obtain the pre-quench single mode Hamiltonians $ \mathcal{H}^{(i)}_k $, with $ i=\{3,4\} $,
\begin{subequations}
	\begin{align}
	\mathcal{H}^{(3)}_{k}&=-2J \cos (k) \sigma^z,\\
	\mathcal{H}^{(4)}_{k}&=-2J(k-\pi/2) \sigma^z,
	\end{align}
\end{subequations}
$ \text{with }k\in[0,\pi) $, for the lattice and the low-energy continuous models, respectively. Also in this model the pre-quench single-mode Hamiltonians are diagonal, and again we can conveniently introduce conduction and valence band Fermi operators using Eq.~\eqref{SM:eq:U0}. The coefficients of the unitary matrix $ U_{0,k}^{(i)} $ are:
\begin{equation}
a^{(i)}_{0,k}=[1-\delta_{k,\pi/2}]\theta\left(\frac{\pi}{2}-k\right)+\frac{\delta_{k,\pi/2}}{\sqrt{2}} \qquad b^{(i)}_{0,k}=[1-\delta_{k,\pi/2}]\theta\left(k-\frac{\pi}{2}\right)+\frac{\delta_{k,\pi/2}}{\sqrt{2}},	\label{SM:eq:ab0SP}
\end{equation}
while the pre-quench conductance and valence bands are
\begin{subequations}
	\begin{align}
	\epsilon_{\pm,0,k}^{(3)}&=\pm 2J|\cos(k)|,\\
	\epsilon_{\pm,0,k}^{(4)}&=\pm 2J |k-\pi/2|.
	\end{align}
\end{subequations}
One obtains $ k_{1}^{(3)}=k_{1}^{(4)}=0 $ and $ k_{2}^{(3)}=k_{2}^{(4)}=\pi $ . Note that we are considering a continuum model with the same number of particles of the lattice one and energy bands filled up to the crossing point. \\

\begin{figure}[h]
	\centering
	\includegraphics[width=0.5\linewidth]{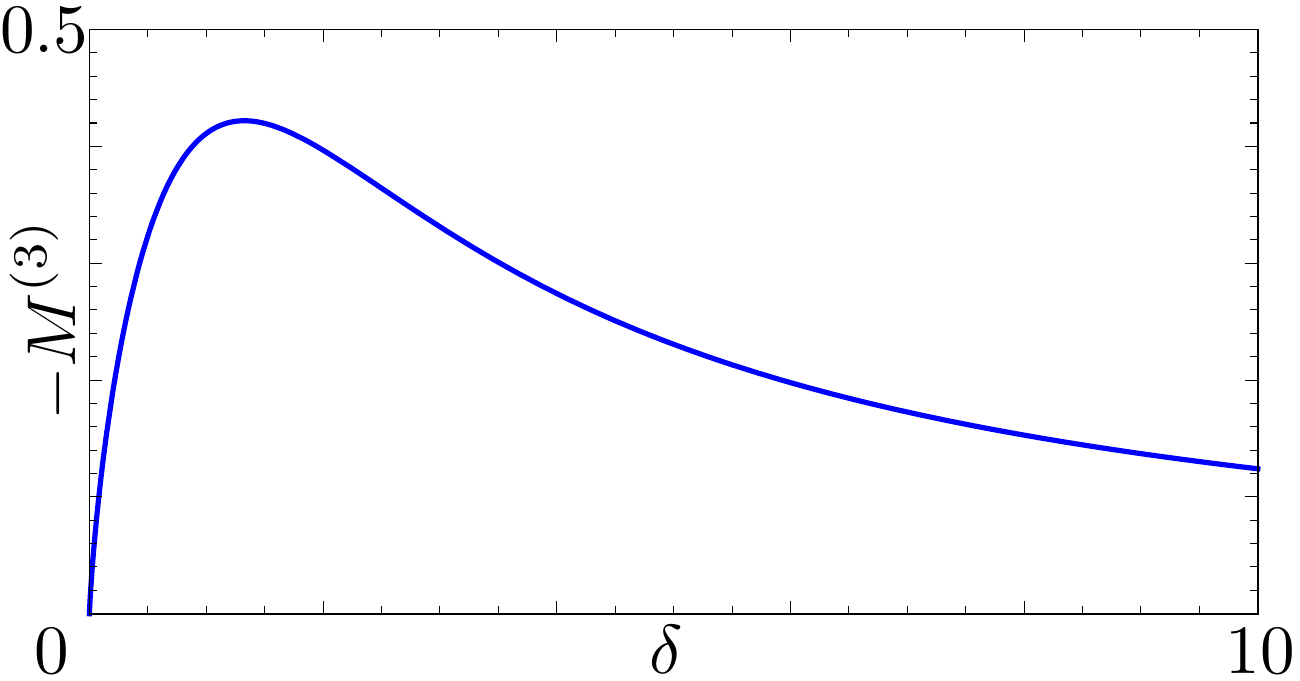}
	\caption{$ M^{(3)} $ for the lattice SP model as a function of $ \Delta^{(3)}=\delta $ and $ J = 1 $.}
	\label{fig:spmodell}
\end{figure}

In the post-quench regime the single-mode Hamiltonian $ \mathcal{H}^{(i)}_k+\delta \sigma^x $ is diagonalized by the unitary matrix $ U^{(i)}_{1,k} $, whose coefficients are
\begin{subequations}
	\label{SM:eq:ab1SP}
	\begin{align}
	a^{(3)}_{1,k}&=\dfrac{\delta}{\sqrt{(\epsilon^{(3)}_{+,1,k}+2J\cos(k))^2+\delta^2}}, &&b^{(3)}_{1,k}=\dfrac{\epsilon^{(3)}_{+,1,k}+2J\cos(k)}{\sqrt{(\epsilon^{(3)}_{+,1,k}+2J\cos(k))^2+\delta^2}},\label{SM:eq:ab1SPL}\\
	a^{(4)}_{1,k}&=\dfrac{\delta}{\sqrt{\left[\epsilon^{(4)}_{+,1,k}+2J(k-\pi/2)\right]^2+\delta^2}}, &&b^{(4)}_{1,k}=\dfrac{\epsilon^{(4)}_{+,1,k}+2J(k-\pi/2)}{\sqrt{\left[\epsilon^{(4)}_{+,1,k}+2J(k-\pi/2)\right]^2+\delta^2}},\label{SM:eq:ab1SPC}
	\end{align}
\end{subequations}
where the new energy bands are 
\begin{subequations}
	\begin{align}
	\epsilon_{\pm,1,k}^{(3)}&=\pm\sqrt{\delta^2+4J^2\cos^2(k)},\\
	\epsilon_{\pm,1,k}^{(4)}&=\pm\sqrt{\delta^2+4J^2(k-\pi/2)^2}.
	\end{align}
\end{subequations}
Using Eqs.~\eqref{SM:eq:M},~\eqref{SM:eq:ab0SP} and~\eqref{SM:eq:ab1SP} one finds that the steady state staggered magnetization after the quench evaluates to 
\begin{equation}\label{SM:eq:MSP}
M^{(i)}=\frac{1}{\pi}\left(\int_{\pi/2}^{\pi}-\int_{0}^{\pi/2}\right)dk\,2a^{(i)}_{1,k}b^{(i)}_{1,k} \left[\left(a^{(i)}_{1,k}\right)^2-\left(b^{(i)}_{1,k}\right)^2\right].
\end{equation}
For the lattice model, using Eq.~\eqref{SM:eq:ab1SPL}, we obtain
\begin{equation}
M^{(3)}=-\dfrac{2\delta}{\pi\sqrt{\delta^2+4J^2}} \text{arctanh}\left(\dfrac{2J}{\sqrt{\delta^2+4J^2}}\right),
\end{equation}
which is shown in Fig.~\ref{fig:spmodell}. On the other hand, from Eq.~\eqref{SM:eq:ab1SPC}, one gets for the low-energy continuous model
\begin{equation}
M^{(4)}=-\dfrac{\delta}{2J\pi}\ln\left[1+\left(\dfrac{\pi J}{\delta}\right)^2\right]
.
\end{equation}

\subsection{2D Rashba-coupled electron gas}
In this Section we consider the quench of magnetic field in a 2D Rashba-coupled electron gas. The Hamiltonian of the system is $ H_{2D}(t)=\sum_{k_x,k_y}\Psi^{\dagger}_{k_x,k_y}[\mathcal{H}_{k_x,k_y}+\theta(t)B\sigma^z]\Psi_{k_x,k_y} $, with 
\begin{equation}
\mathcal{H}_{k_x,k_y}=(k_x^2+k_y^2)I_{2\times 2}+\alpha(\sigma^{x}k_y-\sigma^y{k_x}).
\end{equation}
Here, $ k_x $ and $ k_y $ are the two components of the momentum vector $ \bm{k} $, while $ \Psi^\dagger_{k_x,k_y}=\Big(d^\dagger_{a,k_x,k_y},d^\dagger_{b,k_x,k_y}\Big) $, with $ d_{a,k_x,k_y} $ ($ d_{b,k_x,k_y} $) fermionic annihilation operators for spin up (down) electrons. Following the same steps outlined in the previous Sections, we begin with the pre-quench case. For $ t<0 $ the single-mode Hamiltonian is diagonalized by the unitary matrix
\begin{equation}
U_{0,k_x,k_y}=\begin{bmatrix}
a_{0,k_x,k_y} &b_{0,k_x,k_y}\\
-b^{*}_{0,k_x,k_y} &a_{0,k_x,k_y}
\end{bmatrix},
\end{equation}
with 
\begin{equation}
a_{0,k_x,k_y}=\frac{1}{\sqrt{2}}\qquad b_{0,k_x,k_y}=\frac{1}{\sqrt{2}}\frac{k_x+ik_y}{k},
\end{equation}
where $ k=|\bm{k}|=\sqrt{k_x^2+k_y^2} $. The pre-quench conduction and valence fermionic operators are thus given by
\begin{equation}
\Phi_{0,k_x,k_y}=U_{0,k_x,k_y}\Psi_{k_x,k_y}=\begin{bmatrix}
d_{c,0,k_x,k_y} \\
d_{v,0,k_x,k_y}
\end{bmatrix},
\end{equation}
with associated energy levels
\begin{equation}
\epsilon_{\pm,0,k_x,k_y}=k^2\pm \alpha k.\\
\end{equation}
When the energy bands are filled up to the linear crossing (i.e. the chemical potential is set to zero) the pre-quench equilibrium ground state $ |\Phi^{(2D)}_{0}(0)\rangle $ is
\begin{equation}
|\Phi^{(2D)}_{0}\rangle=\prod_{k\leq\alpha}\left(\Phi^{\dagger}_{0,k_x,k_y}\right)_2|0_{2D}\rangle=\prod_{k\leq\alpha}\left(U^{\dag}_{0,k_x,k_y}\Psi^{\dagger}_{k_x,k_y}\right)_2|0_{2D}\rangle,
\end{equation}
with $ |0_{2D}\rangle $ the vacuum of the system. As usual, the subscript $ 2 $ means that the second component of the spinor has to be considered. 

\begin{figure}[h]
	\centering
	\includegraphics[width=0.5\linewidth]{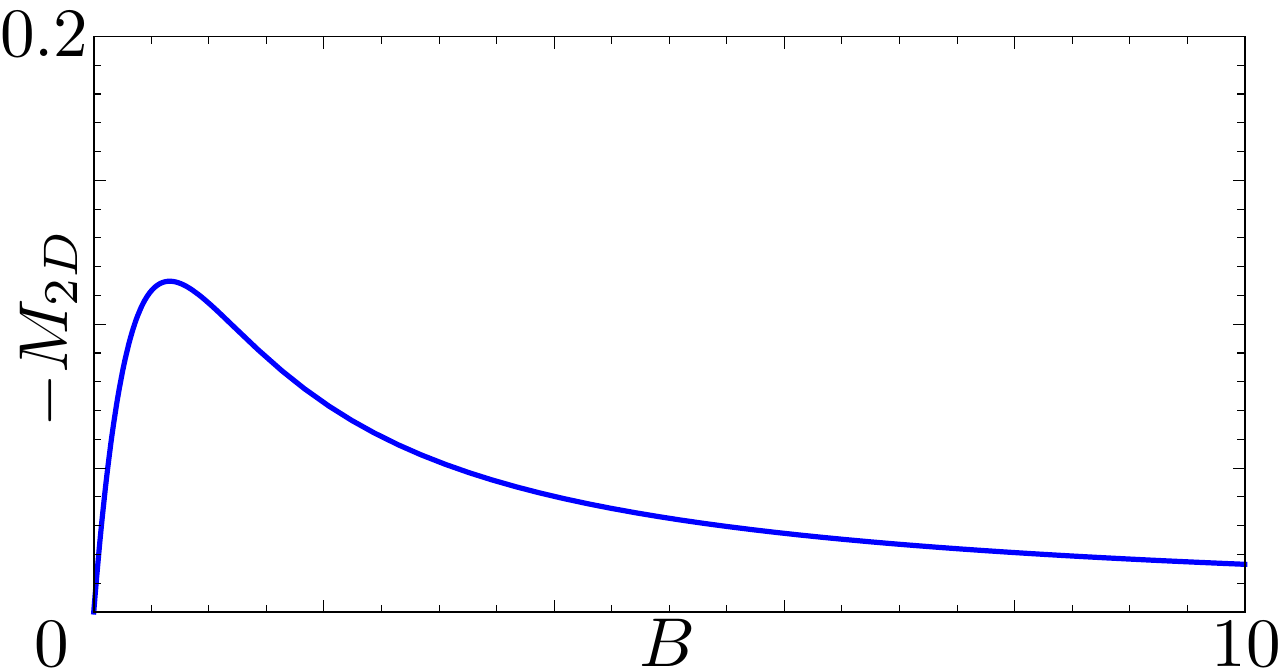}
	\caption{$ M_{2D} $ for the 2D Rashba-coupled electron gas as a function of $ B $ and $ \alpha = 1 $.}
	\label{fig:socwire2d}
\end{figure}

We now turn to the post-quench regime. For $ t>0 $ the unitary matrix diagonalizing the single-mode Hamiltonian $\mathcal{H}_{k_x,k_y}+B\sigma^z $ is 
\begin{equation}
U_{1,k_x,k_y}=\begin{bmatrix}
a_{1,k_x,k_y} &b_{1,k_x,k_y}\\
-b^{*}_{1,k_x,k_y} &a_{1,k_x,k_y}
\end{bmatrix},
\end{equation}
with 
\begin{equation}
a_{1,k_x,k_y}=\frac{\alpha k} {\sqrt{(D_{k_x,k_y}-B)^2+\alpha^2 k^2}},\qquad b_{1,k_x,k_y}=\frac{D_{k_x,k_y}-B}{\sqrt{(D_{k_x,k_y}-B)^2+\alpha^2 k^2}}\frac{k_x+ik_y}{k},
\end{equation}
where we have introduced the coefficient $ D_{k_x,k_y}=\sqrt{B^2+\alpha^2k^2}$. The post-quench conductance and valence band Fermi operators are 
\begin{equation}
\Phi_{1,k_x,k_y}=U_{1,k_x,k_y}\Psi_{k_x,k_y}=\begin{bmatrix}
d_{c,1,k_x,k_y} \\
d_{v,1,k_x,k_y}
\end{bmatrix},
\end{equation}
with associated energy levels
\begin{equation}
\varepsilon_{\pm,1,k_x,k_y}=(k_x^2+k_y^2)\pm D_{k_x,k_y}.\\
\end{equation}
In order to get the steady state magnetization along the applied magnetic field within the GGE picture, we evaluate the averages of the conserved occupation numbers of the post-quench energy levels,
\begin{equation}
n_{k_x,k_y,j=1,2}=\left(\Psi^\dagger_{k_x,k_y}U^{\dagger}_{1,k_x,k_y}\right)_j \left( U_{1,k_x,k_y}\Psi_{k_x,k_y}\right)_j,
\end{equation}
over the pre-quench ground state $ |\Phi^{(2D)}_{0}(0)\rangle $, obtaining
\begin{subequations}
	\begin{align}
	\langle n_{k_x,k_y,1}\rangle_0&=\langle n_{k_x,k_y,1}\rangle_{GGE}=\left|-a_{1,k_x,k_y}b_{0,k_x,k_y}+a_{0,k_x,k_y}b_{1,k_x,k_y}\right|^2\langle d^{\dagger}_{v,0,k_x,k_y} d_{v,0,k_x,k_y}\rangle_0,\\
	\langle n_{k_x,k_y,2}\rangle_0&=\langle n_{k_x,k_y,2}\rangle_{GGE}=\left|a_{1,k_x,k_y}a_{0,k_x,k_y}+b_{0,k_x,k_y}b^{*}_{1,k_x,k_y}\right|^2\langle d^{\dagger}_{v,0,k_x,k_y} d_{v,0,k_x,k_y}\rangle_0.
	\end{align}
\end{subequations}
Since $ \langle d^\dagger_{c,1,k_x,k_y}d_{v,1,k_x,k_y}\rangle_{GGE}=\langle d^\dagger_{v,1,k_x,k_y}d_{c,1,k_x,k_y}\rangle_{GGE}=0 $, the steady state magnetization after the quench evaluates to
\begin{align}
M_{2D}&=\frac{1}{N_{2D}}\sum_{k_x,k_y}\langle\Psi^\dagger_{k_x,k_y}\sigma^z\Psi_{k_x,k_y}\rangle_{GGE}\\
&=\frac{1}{N_{2D}}\sum_{k_x,k_y}\left(a_{1,k_x,k_y}^2-|b_{1,k_x,k_y}|^2\right)\left(\langle n_{k_x,k_y,1}\rangle_{GGE}-\langle n_{k_x,k_y,2}\rangle_{GGE}\right)\\
&=-\frac{B}{\alpha^2}\left[1-\frac{B}{\alpha^2}\text{arccot}\left(\frac{B}{\alpha^2}\right)\right],
\end{align}
where in the last step the thermodynamic limit has been performed and we used that $ N_{2D}=L_xL_y/(2\pi)^2 $, with $ L_x $ and $ L_y $ the length of the system in the $ x $ and $ y $ directions respectively. The behavior of $ M_{2D} $ is shown in Fig.~\ref{fig:socwire2d}.

\section{Klein-Gordon physics in the evolution of the Green's function}
In this Section we focus on the low-energy continuous theories for the SOC wire and the SP model and derive KG equation [see Eq.~(7) of the main text] satisfied by the general Green's function
\begin{equation}
\mathcal G^{(i)}(x,t)=\langle\Psi^{(i)\dagger}(x,t)\sigma^{x}\Psi^{(i)}(0,t)\rangle_0\label{SM:eq:G}
\end{equation}
after a sudden quench of the gap opening mechanism. Here, $ \Psi^{(i)\dag}(x,t)=\left(\psi^{(i)\dag}_a(x,t),\psi^{(i)\dag}_b(x,t)\right) $ is the space-resolved Fermi spinor in the Heisenberg picture, while the average is evaluated on the pre-quench equilibrium ground state $ |\Phi_0^{(i)}\rangle $.
\subsection{Spin-orbit coupled wire}
We start by deriving the equation of motion of the Fermi field operator $ \Psi^{(2)}(x,t) $ for the SOC wire. By rewriting the Hamiltonian of Eq.~\eqref{SM:eq:H} in real space as a function of $ \Psi^{(2)}(x) $ one gets the Heisenberg equations of motion for the Fermi spinor components,
\begin{equation}
\partial_t \psi_\sigma^{(2)}(x,t)=\left(i\partial_x^2-\sigma\alpha\partial_x\right)\psi_\sigma^{(2)}(x,t)-iB\psi_{-\sigma}^{(2)}(x,t),
\end{equation}
with $ \sigma,\sigma'=\{a,b\}=\{+,-\} $. From the above equation, we can derive the equations of motion for the spin resolved Green's functions,
\begin{equation}
G^{(i)}_{\sigma\sigma'}(x,t)=\langle \psi^{(i)\dag}_{\sigma}(x,t) \psi^{(i)}_{\sigma'}(0,t)\rangle_0.
\end{equation}
As a result, we obtain the following closed set of differential equations
\begin{subequations}
	\label{SM:eq:closedGs}
	\begin{align}
	\partial_t G^{(2)}_{++}(x,t)&=-iB\left[G^{(2)}_{+-}(x,t)-G^{(2)}_{-+}(x,t)\right],\label{eq:g++}\\
	\partial_t G^{(2)}_{+-}(x,t)&=-iB\left[G^{(2)}_{++}(x,t)-G^{(2)}_{--}(x,t)\right]+2\alpha\partial_x G^{(2)}_{+-}(x,t),\label{eq:g+-}\\
	\partial_t
	G^{(2)}_{-+}(x,t)&=+iB\left[G^{(2)}_{++}(x,t)-G^{(2)}_{--}(x,t)\right]-2\alpha\partial_x G^{(2)}_{-+}(x,t),\label{eq:g-+}\\
	\partial_t
	G^{(2)}_{--}(x,t)&=+iB\left[G^{(2)}_{+-}(x,t)-G^{(2)}_{-+}(x,t)\right].\label{eq:g--}
	\end{align}
\end{subequations}
Note that the Green's function of Eq.~\eqref{SM:eq:G} can be written as $ \mathcal G^{(2)}(x,t)=G^{(2)}_{+-}(x,t)+G^{(2)}_{-+}(x,t) $. From Eqs.~\eqref{SM:eq:closedGs}, we thus obtain
\begin{equation}
\partial_t^2 \mathcal G^{(2)}(x,t)=2\alpha\partial_x \partial_t\left[G^{(2)}_{+-}(x,t)-G^{(2)}_{-+}(x,t)\right]=4\alpha^2\partial_x^2 \mathcal G^{(2)}(x,t) -4i\alpha B\partial_x\left[G^{(2)}_{++}(x,t)-G^{(2)}_{--}(x,t)\right].
\label{SM:eq:EoMG1}
\end{equation}
It is now convenient to introduce the function $ S(x,t)=G^{(2)}_{++}(x,t)-G^{(2)}_{--}(x,t) $, whose time derivative reads
\begin{equation}
\partial_t S(x,t)=-2iB\left[G^{(2)}_{+-}(x,t)-G^{(2)}_{-+}(x,t)\right].
\end{equation}
By integrating the above equation, one obtains
\begin{equation}\label{eq:s}
S(x,t)=S(x,0)-2iB\int_{0}^{t}\left[G^{(2)}_{+-}(x,t')-G^{(2)}_{-+}(x,t')\right]dt'.
\end{equation}
Then, by taking the space derivative of $ S(x,t) $ and noting that  $ \partial_x \left[G^{(2)}_{+-}(x,t')-G^{(2)}_{-+}(x,t')\right] =(2\alpha)^{-1}\partial_t \mathcal G^{(2)}(x,t) $ [see Eqs.~\eqref{eq:g+-} and~\eqref{eq:g-+}], it follows that
\begin{equation}
\partial_x S(x,t)=\partial_x S(x,0)-i\dfrac{B}{\alpha} \mathcal G^{(2)}(x,t).
\end{equation}
Finally, turning back to Eq.~\eqref{SM:eq:EoMG1}, we obtain the desired result
\begin{equation}
\left(\partial_x^2-\frac{1}{4\alpha^2}\partial_t^2\right) \mathcal G^{(2)}(x,t) = \frac{B^2}{\alpha^2}  \mathcal G^{(2)}(x,t)+\frac{B}{\alpha}\phi_2(x),
\label{SM:eq:EoMG2}
\end{equation}
where the source term $ \phi_2(x) $ is defined as
\begin{equation}
\phi_2(x)=i\partial_x S(x,0)=i\partial_x \langle\Psi^{(2)\dagger}(x,0)\sigma^z\Psi^{(2)}(0,0) \rangle_0.
\end{equation}
In particular, $ \phi_{2}(x) $ can be analytically evaluated to obtain
\begin{subequations}
	\begin{align}
	\phi_2(x)&=2\left[\dfrac{1-\cos(\alpha x)}{x^2}-\dfrac{\alpha\sin(\alpha x)}{x}\right].
	\end{align}
\end{subequations}
\subsection{Staggered potential model}
We now focus on the SP model. In principle, the KG equation satisfied by the Green's function $ \mathcal G^{(4)}(x,t) $ can be obtained following the same steps of the SOC wire case. However, in order to show an alternative method to derive it, we demonstrate that $ \mathcal G^{(4)}(x,t) $ satisfies the analogous of Eq.~\eqref{SM:eq:EoMG2} by a direct calculation. We begin by explicitly evaluating $ \mathcal G^{(4)}(x,t)=\langle\Psi^{(4)\dag}(x,t)\sigma^x\Psi^{(4)}(0,t)\rangle $. The time evolution of the Fermi spinor $ \Psi^{(4)}(x,t)=\sum_k \Psi_k^{(4)}(t) e^{ikx}/\sqrt{L^{(4)}}$ in the Heisenberg picture can be obtained from Eq.~(2) of the main text,
\begin{equation}\label{eq:tev4}
\Psi_k^{(4)}(t)= U_{1,k}^{(4)\dag}\mathrm{diag}\{e^{-i\epsilon_{+,1,k}^{(4)}t},e^{-i\epsilon_{-,1,k}^{(4)}t}\}U_{1,k}^{(4)} U_{0,k}^{(4)\dag} \Phi_{0,k}^{(4)}(0),
\end{equation}
with the coefficients of the matrices $ U^{(4)}_{0,k} $ and $ U^{(4)}_{1,k} $ given in Eqs.~\eqref{SM:eq:ab0SP} and \eqref{SM:eq:ab1SPC}, respectively. Here, $ \Psi^{(i)\dagger}_k(t)=\left(d_{a,k}^{(i)\dagger}(t),d_{b,k}^{(i)\dagger}(t)\right) $ is the momentum resolved Fermi spinor and $ L^{(4)} $ is the length of the system. The Green's function $ G^{(4)}(x,t) $ can thus be rewritten as
\begin{equation}
\mathcal G^{(4)}(x,t)=\frac{1}{L^{(4)}}\sum_k e^{-ikx}\langle d^{(4)\dag}_{b,k}(t)d^{(4)}_{a,k}(t)+h.c.\rangle_0, 
\label{SM:eq:G4}
\end{equation}
where the average is evaluated on the ground state of the pre-quench Hamiltonian $ \mathcal{H}_k^{(4)} $, defined in Eq.~\eqref{SM:eq:GS0}.
Using Eqs.~\eqref{SM:eq:ab0SP},~\eqref{SM:eq:ab1SPC} and \eqref{eq:tev4}, we obtain
\begin{equation}
\langle d^{(4)\dag}_{b,k}(t)d^{(4)}_{a,k}(t) \rangle_0 = \frac{1}{8} \langle d^{(4)\dag}_{v,1,k} d^{(4)}_{v,1,k}\rangle_0 \left[-4\beta_k\text{Im}\{\beta_k\}-ie^{-2it\epsilon^{(4)}_{+,1,k}}\left(1+\beta_k^{2}\right)-ie^{-2it\epsilon^{(4)}_{-,1,k}}\beta_k^{2}\left(1+\beta_k^{*2}\right)\right],
\end{equation}
where $ \beta_k=\sqrt{2}b^{(4)}_{1,k} $. Substituting in Eq.~\eqref{SM:eq:G4} and performing the thermodynamic limit, one has
\begin{equation}
\mathcal G^{(4)}(x,t)=- \frac{1}{\pi} \int_{0}^{\pi} e^{-ikx}\dfrac{J k\delta}{J^2k^2+\delta^2}\left[1-\cos\left(2t\epsilon_{+,1,k}^{(4)}\right)\right]dk.
\end{equation}
Finally, after evaluating the second-order time and space derivatives of $ G^{(4)}(x,t) $,
\begin{subequations}
	\begin{align}
	&\partial_t^2 \mathcal G^{(4)}(x,t)=-\frac{4}{\pi} \int_{0}^{\pi}e^{-ikx}J k\delta\cos\left(2t\epsilon_{+,1,k}^{(4)}\right)dk,\\
	&\partial_x^2 \mathcal G^{(4)}(x,t)=\frac{1}{\pi} \int_{0}^{\pi}e^{-ikx}\dfrac{J k^3\delta}{J^2k^2+\delta^2}\left[1-\cos\left(2t\epsilon_{+,1,k}^{(4)}\right)\right]dk,
	\end{align}
\end{subequations}
and performing  some algebraic manipulations, one can directly verify that the following KG equation is satisfied
\begin{equation}
\left(\partial_x^2-\frac{1}{4J^2}\partial_t^2\right) \mathcal G^{(4)}(x,t) = \frac{\delta^2}{J^2}  \mathcal G^{(4)}(x,t)+\frac{\delta}{J}\phi_4(x),
\end{equation}
where the source term $ \phi_4(x) $ is
\begin{equation}
\phi_4(x)=2\frac{\cos(\pi x)+\pi x \sin(\pi x)-1}{\pi x^2}=i\partial_x \langle \Psi^{(4)\dag}(x,0)\sigma^{y}\Psi^{(4)}(0,0) \rangle_0.
\end{equation}

\section{Finite duration quench for the spin-orbit coupled wire}
In this last Section we outline the evaluation of the steady state magnetization of the SOC wire in the presence of a quench with finite duration. In particular, we consider a quench protocol in which the magnetic field is switched on with a linear ramp of duration $ \tau $. The Hamiltonian of the systems is
\begin{equation}
H=\sum_k \Psi_k^{(2)\dag}\left[\mathcal{H}^{(2)}_k+Q(t)B\sigma^x\right]\Psi_k^{(2)},
\end{equation}
with
\begin{equation}
\label{SM:eq:Q}
Q(t)=\left\{\begin{array}{l}
0\\ 
t/\tau\\ 
1
\end{array} \right.
\begin{array}{l}
\text{for } t<0\\ 
\text{for } 0\leq t\leq\tau\\
\text{for } t> \tau 
\end{array}.
\end{equation}
During and after the quench, the Heisenberg equations of motion for the Fermi spinor components are
\begin{equation}
\label{SM:eq:EoMd}
\partial_t d^{(2)}_{\sigma,k}(t)=-i\left[(k^2+\sigma\alpha k)d^{(2)}_{\sigma,k}(t)+Q(t)Bd^{(2)}_{-\sigma,k}(t)\right],
\end{equation}
where $ \sigma=\{a,b\}=\{+,-\} $. To solve this coupled system of differential equations, we take the following ansatz~\cite{Dora:2011}
\begin{equation}
\label{SM:eq:ansatz}
\left[\begin{array}{c}
d^{(2)}_{a,k}(t)\\ 
d^{(2)}_{b,k}(t)
\end{array} \right]=\left[\begin{array}{cc}
f_{a,k}(t) & g_{a,k}(t) \\ 
f_{b,k}(t) & g_{b,k}(t)
\end{array} \right] \left[\begin{array}{c}
d^{(2)}_{a,k}\\ 
d^{(2)}_{b,k}
\end{array} \right]=V_k(t)\left[\begin{array}{c}
d^{(2)}_{a,k}\\ 
d^{(2)}_{b,k}
\end{array} \right],
\end{equation}
where $ d^{(2)}_{\sigma,k} $ is the Fermi operator in the Schr\"odinger picture at $ t=0 $. Therefore, all the time dependence is encoded in the functions $ f_{\sigma,k}(t) $ and $ g_{\sigma,k}(t) $, with initial conditions given by $ f_{a,k}(0)=g_{b,k}(0)=1 $ and $ f_{b,k}(0)=g_{a,k}(0)=0 $. Since anti-commutation relations between the operators $ d^{(2)}_{\sigma,k}(t) $  have to be satisfied during the whole time evolution,we have that $ |f_{\sigma,k}(t)|^2+|g_{\sigma,k}(t)|^2=1,\ \forall t $. By substituting the ansatz of Eq.~\eqref{SM:eq:ansatz} in Eq.~\eqref{SM:eq:EoMd}, we obtain two decoupled systems for $ f_{\sigma,k}(t) $ and $ g_{\sigma,k}(t) $, respectively,
\begin{equation}
\label{SM:eq:decoupledsystems}
i\partial_t \left[\begin{array}{l}
f_{a,k}(t)\\ 
f_{b,k}(t)
\end{array} \right]=\left[\begin{array}{cc}
k^2+\alpha k & Q(t)B \\ 
Q(t)B & k^2-\alpha k
\end{array} \right]\left[\begin{array}{l}
f_{a,k}(t)\\ 
f_{b,k}(t)
\end{array} \right] \qquad i\partial_t \left[\begin{array}{l}
g_{a,k}(t)\\ 
g_{b,k}(t)
\end{array} \right]=\left[\begin{array}{cc}
k^2+\alpha k & Q(t)B \\ 
Q(t)B & k^2-\alpha k
\end{array} \right]\left[\begin{array}{l}
g_{a,k}(t)\\ 
g_{b,k}(t)
\end{array} \right].
\end{equation}
The latter systems can be solved with same method, given that the appropriate initial conditions are used. In particular, introducing the notation $ \nu=\{f,g\} $, we define the functions~\cite{Porta:2016}
\begin{equation}
S_{\nu,k}(t)=\nu_{a,k}(t)+\nu_{b,k}(t), \qquad D_{\nu,k}(t)= \nu_{a,k}(t)-\nu_{b,k}(t).
\end{equation}
Using Eq.~\eqref{SM:eq:decoupledsystems}, one obtains that the following differential equations hold
\begin{equation}\label{eq:sysSD}
\left\{\begin{array}{l}
i\partial_t S_{\nu,k}(t) = \left[k^2+Q(t)B\right]S_{\nu,k}(t)+\alpha k D_{\nu,k}(t)\\ 
i\partial_t D_{\nu,k}(t) = \left[k^2-Q(t)B\right]D_{\nu,k}(t)+\alpha k S_{\nu,k}(t)
\end{array} \right. ,
\end{equation}
From the above system we derive the second-order differential equation
\begin{equation}\label{eq:eqdiffD}
\partial^2_t D_{\nu,k}(t)+2ik^2 \partial_t D_{\nu,k}(t)+\left[B^2Q^2(t)-k^4+\alpha^2 k^2-iB\partial_t Q(t) \right]D_{\nu,k}(t)=0,
\end{equation}
which can be analytically solved in every region defined by the quench protocol in Eq.~\eqref{SM:eq:Q} using the appropriate matching conditions on the boundaries of each them. Moreover, once we get $ D_{\nu,k}(t) $, the function $ S_{\nu,k}(t) $ is automatically determined by the second equation in Eq.~\eqref{eq:sysSD}.\\

\begin{figure}[h]
	\centering
	\includegraphics[width=0.5\linewidth]{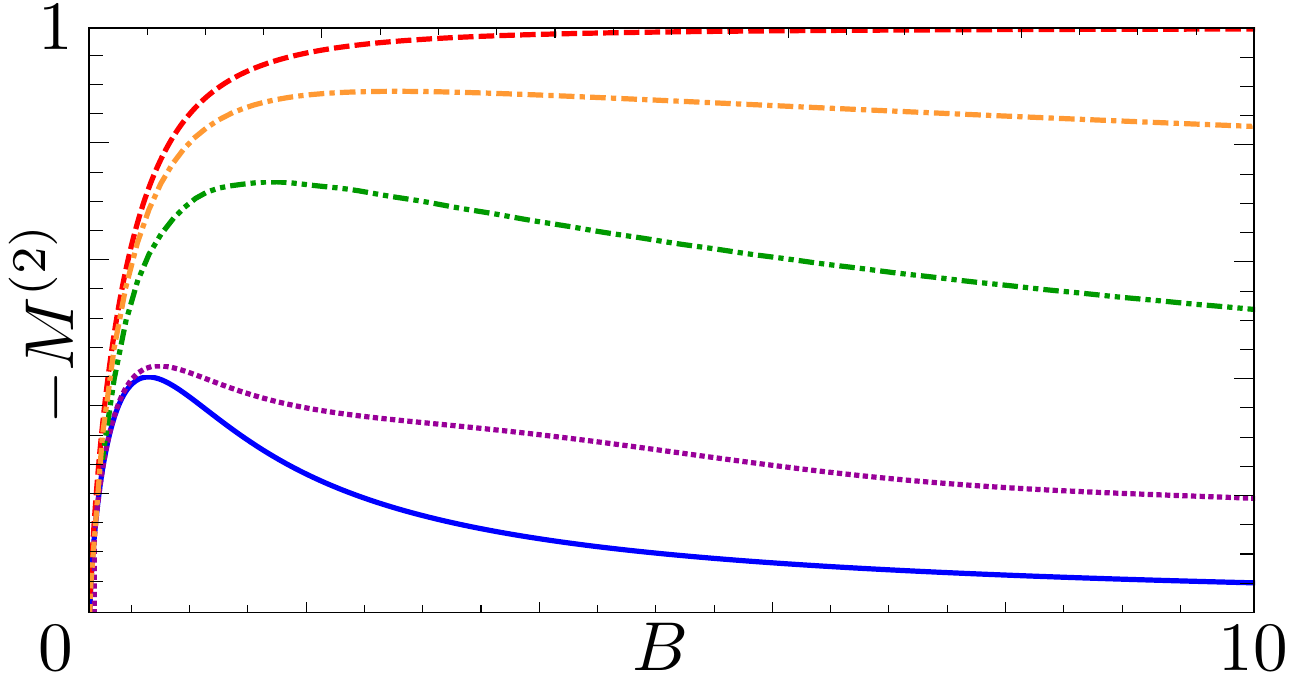}
	\caption{$ M^{(2)} $ for the lattice SOC wire as a function of $ \Delta^{(1)}=B $ and $ \alpha = 1 $ for different quench protocols with increasingly long switch-on times $ \tau $ (units $ 1/\alpha^2 $): $ \tau=0 $, sudden quench (blue, solid), $ \tau=1 $ (purple, dot), $ \tau=10 $ (green, dash-dot-dot), $ \tau=100 $ (yellow, dash-dot), $ \tau=\infty $ -corresponding
		to the equilibrium magnetization of the post-quench Hamiltonian
		red - (red, dash).
	}
	\label{fig:soctau}
\end{figure}

The magnetization along the applied magnetic field can be evaluated within the GGE, with a straightforward generalization of procedure described in Sec.~\ref{SM:sec:generalM}. In particular, the quantities conserved after the quench (i.e. for $ t>\tau $) are $ \langle n_{k,j}^{(2)}(\tau)\rangle_0=\langle n_{k,j}^{(2)}(\tau)\rangle_{GGE} $, with
\begin{equation}
n_{k,j}^{(2)}(\tau)= \left(\Phi_{0,k}^{(2)\dag}U_{0,k}^{(2)}V^\dag_k(\tau)U_{1,k}^{(2)\dag}\right)_j \left(U_{1,k}^{(2)}V_k(\tau)U_{0,k}^{(2)\dag}\Phi_{0,k}^{(2)}\right)_j
\end{equation}
the occupation numbers of the post-quench energy levels and the unitary matrix $ V_k(t) $ introduced in Eq.~\eqref{SM:eq:ansatz}. From the knowledge of $\langle n_{k,j}^{(2)}(\tau)\rangle_{GGE} $ and thanks to the fact that $ \langle d^{(i)\dagger}_{c,1,k}(\tau)d^{(i)}_{v,1,k}(\tau)\rangle_{GGE}=\langle d^{(i)\dagger}_{v,1,k}(\tau)d^{(i)}_{c,1,k}(\tau)\rangle_{GGE}=0 $,
one can evaluate the steady state magnetization
\begin{equation}
M^{(2)}=\frac{1}{n^{(2)}}\sum_k \langle\Psi^{(2)\dagger}_k\sigma^x\Psi^{(2)}_k\rangle_{GGE}=\frac{1}{n^{(2)}}\sum_{k} \langle \Phi_{1,k}^{(2)\dag}(\tau)U_{1,k}^{(2)}V_k(\tau)\sigma^x V^\dag_k(\tau) U_{1,k}^{(2)\dag} \Phi_{1,k}^{(2)}(\tau) \rangle_{GGE}.
\end{equation}
The behavior of $ M^{(2)} $ for quench protocols with different time duration $\tau $ is shown in Fig.~\ref{fig:soctau}. As one can clearly see, the non-monotonic behavior persists also in this case.

\section{Interaction quench}
In this Section we give more details about the interaction quench discussed in the main paper. The real space lattice Hamiltonian is defined by
\begin{equation}
H=-J\sum_{j} \left[\left(c^\dag_jc_{j+1}+h.c.\right)+U(t)n_jn_{j+1}+V(t)n_jn_{j+2}\right],
\end{equation}
where $ c_j $ are annihilation operators of spinless fermions on the site $ j $ of the lattice and $n_j=c^\dag_jc_{j}$ is the density. The Hamiltonian is thus similar to the one of Eq.~\eqref{SM:eq:H_SP}, but instead of quenching a staggered potential, here a nearest and possibly next-nearest neighbor interaction is quenched on by $U(t)=U\theta(t)$ and $V(t)=V\theta(t)$. For $V=0$ the system is Bethe-Ansatz solvable (integrable) while for $V\neq0$ no solution via the Bethe-Ansatz is known (and is believed to be nonexistent). We consider quenches out of the initial state  $\left|E_0\right>$ for $U=V=0$ and perform  the time evolution w.r.t. a finite $U$ and possibly $V$. At $V=0$ it is known, that for $U/J<2$ the system remains gapless after the quench, while for $U/J>2$ a gap opens. 

In contrast to the examples studied above, this interacting Hamiltonian cannot be diagonalized analytically and we resort to numerical means. We employ the density matrix renormalization group (DMRG) \cite{White92,Schollwoeck11}. In practice the accuracy of DMRG simulations is controlled by the so-called 
bond dimension $\chi$. By increasing $\chi$ (and with it the numerical effort) we
achieve converged results, which are ``numerically exact". In every simulation we choose $\chi$ such that no changes 
of the results can be observed on the scales of the respective plots if it is further increased. 

To obtain numerical results we need to
\begin{enumerate}
	\item prepare the ground state of the noninteracting ($U=0=V$) system,
	\item perform the time evolution with respect to finite $U$ and possibly $V$,
	\item calculating the finite temperature canonical ensemble (as a reference), w.r.t. finite $U$ and possibly $V$, 
\end{enumerate}
the third of which is independent of the former two.

All of these are routinely coded very elegantly in the limit of infinite lattice size 
$n \to \infty$ \cite{Vidal07,Orus08}  using the language 
of matrix product states (MPS) \cite{Schollwoeck11}. For more implementation details we refer to the aforementioned references. The ground state is obtained here by an imaginary time evolution. Starting with a random MPS and repetitively applying $e^{-H(t<0)\Delta\tau}$ (normalizing the MPS after each step), we converge to the ground state of the system by projecting out excited states. We apply $e^{-H(t<0)\Delta\tau}$ by employing a second order Trotter decomposition. 
We start with larger values of $\Delta\tau$ and successively lower it until the MPS converges, where we aim at a relative accuracy of the total energy per lattice 
site of $10^{-10}$. As the ground state we start with is the one of the non-interacting system it is easy to benchmark the accuracy of the found ground state by comparing to exact results. The subsequent real-time evolution of this MPS is performed  by 
repetitively applying $e^{-iH(t>0)\Delta t}$, using a fourth order Suzuki-Trotter decomposition and choosing $\Delta t/J=0.1$ small enough such 
that the error arising from this decomposition is negligible. In a similar fashion we determine the finite temperature canonical ensemble. We use purification at $\beta=1/T=0$ effectively rewriting the ensemble as an 
MPS in enlarged physical space \cite{Verstraete04}. We then  ``cool down"
the MPS by repetitively applying $e^{-H(t>0)\Delta\beta}$ (and normalizing) in an 
appropriately Trotter decomposed fashion. 

Within the DMRG we have access to finite times only (as the results are obtained by explicit forward time evolution). To push the simulations to larger time scales  we employ the ideas of Ref.~\cite{Kennes16}. Of course (like in any finite time simulation) we cannot exclude that the results obtained are not truly steady and that on some very large time scale another relaxation mechanism sets in. However, we can access time scales on which typical observables appear approximately relaxed on the scale of their respective plots. We might thus only have access to the prethermal state of the system, stipulating that on inaccessible timescales a second regime shows up, which changes the steady state value. Either way the results reported here are then (at least) valid on an extensively long time scales, which would be relevant to experiments.


\begin{thebibliography}{99}
\bibitem{footnote:continuumSSH} Note that in order to mimic the finite energy range we have set the summation range between $ -\pi $ and $ \pi $ also in the continuum model.
\end{thebibliography}
\end{document}